# New Galactic $\beta$ Lyrae-type Binaries Showing Superorbital Photometric Cycles

Gonzalo Rojas García[1,2], Ronald Mennickent[2], P. Iwanek[3], P. Gorrini[2], J. Garcés[2], I. Soszyński[3], and N. Astudillo-Defru[4]
[1] Centrum Astronomiczne im. Mikołaja Kopernika (CAMK), PAN, Bartycka 18, 00-716 Warsaw, Poland; gonzrojas@udec.cl
[2] Universidad de Concepción, Departamento de Astronomía, Casilla 160-C, Concepción, Chile
[3] Astronomical Observatory, University of Warsaw, Al. Ujazdowskie 4, PL-00-478 Warszawa, Poland
[4] Departamento de Matemática y Física Aplicadas, Universidad Católica de la de Santísima Concepción, Alonso de Rivera 2850, Concepción, Chile
*Received 2021 March 15; revised 2021 July 22; accepted 2021 July 24; published 2021 November 16*

## Abstract

We present the discovery of 32 new double periodic variables (DPVs) located toward the Galactic bulge. We found these objects among the nearly half a million binary stars published by the Optical Gravitational Lensing Experiment project. With this discovery, we increase the number of known DPVs in the Milky Way by a factor of 2. The new set of DPVs contains 31 eclipsing binaries and one ellipsoidal variable star. The orbital periods cover the range from 1.6 to 26 days, while long periods are detected between 47 and 1144 days. Our analysis confirms a known correlation between orbital and long periods that is also observed in similar systems in the Magellanic Clouds.

*Unified Astronomy Thesaurus concepts:* Eclipsing binary stars (444); Beta Lyrae stars (149); Double periodic variable stars (2111)

## 1. Introduction

Double periodic variables (DPVs) are a class of semidetached binary systems that have the remarkable characteristic of a secondary long photometric variation in their light curves in addition to the variation due the orbital motion. This class of binaries has been noticed for the first time by Mennickent et al. (2003) after a search for Be-type stars in the Magellanic Clouds. Later, many more similar variables were found in the Magellanic Clouds (Poleski et al. 2010; Pawlak et al. 2013). Previously, some Algols were reported in the literature to have a long periodicity. Kalv (1979) interpreted the 516 day periodicity of RX Cas as pulsations of the Roche-lobe-filling star. Pulsations for the B8 II component of $\beta$ Lyrae are also suggested by Guinan (1989), although he also mentions possible changes in the structure of the disk. Peters (1994) suggests that the AU Mon long-term light variability is caused by cyclic pulsations of the donor. Studying $\beta$ Lyrae, Harmanec et al. (1996) considered the 282 day cycle as a possible beat between the 12.9 day orbital period and the 4.7 day period detected in spectroscopy.

Semidetached systems, in particular DPVs, are interacting binaries where mass transfer occurs between the stellar components. The gravitational potential in this configuration can be expressed by the Roche potential (Roche 1873; Kopal 1959), where one of the components (cooler and less massive, the secondary component) overflows its Roche lobe, transferring mass onto the primary component (hotter and more massive, the primary component). As a consequence of the mass transfer, an accretion disk can appear surrounding the primary component. An extensive overview of DPVs has been published by Mennickent (2017). DPVs generally host B-type dwarf primary components of mass between 7 and 10 $M_\odot$ and A/F/G-type giant secondary components with masses between 1–3 $M_\odot$. The intrinsic relation between the orbital and long periods can be expressed by $P_l \approx 33 P_{\rm orb}$, although single systems can shift from this general tendency appreciably (Mennickent 2017). The origin of the long cycle still is an enigma, although Schleicher & Mennickent (2017) argued that a magnetic dynamo in the donor drives the long cycle. As explained by Mennickent et al. (2016), the physical configuration of these sources is quite similar to those found in classical Algols, but DPVs are hotter and more massive.

From the evolutionary point of view, DPVs can be considered Algols because of their intermediate mass and reversal of the mass ratio, with the less massive component being the more evolved one (Crawford 1955). This in principle puzzling condition was once called "the Algol paradox." Nowadays, it is unknown why only some Algols display the DPV phenomenon, and is clear that the discovery of new systems will help to better understand the physical processes in close-in binary systems and explain the origin of their superorbital cycles.

Only a few DPVs have been discovered in the Milky Way, compared with significantly larger numbers reported in the Magellanic Clouds (Mennickent et al. 2003; Poleski et al. 2010; Pawlak et al. 2013). It is then expected that several Galactic DPVs remain to be be discovered in our galaxy. The context of some photometric surveys still need to be searched for DPVs, e.g., Kepler (Kirk et al. 2016) and the Catalina Survey (Drake et al. 2009). In order to start filling this gap, we searched for new DPVs in the catalog of eclipsing and ellipsoidal binary stars toward the Galactic bulge discovered in the framework of the Optical Gravitational Lensing Experiment (OGLE) project (Soszyński et al. 2016). Here we present the results of this inspection.

## 2. Observation and Data

We searched for new DPVs in the catalog of more than 450,000 eclipsing and ellipsoidal binary stars published by Soszyński et al. (2016), considering data obtained under phases II, III, and IV of the OGLE project. The data were obtained at the 1.3 m Warsaw Telescope, located at Las Campanas Observatory, Chile, during the years 1997–2015. Images were taken in two filters, $V$ and $I$. The data reduction and calibrations are described in Udalski et al. (2015). Since 2010 until today the OGLE-IV project regularly has observed about 400 million stars in 182 deg$^2$ of the densest regions of the Galactic bulge. The search for eclipsing variables was based primarily on the OGLE-IV data. The catalog includes objects with magnitudes $12 < I < 21$ mag and $13 < V < 21.7$ mag. The CCD saturates at about 13 mag in the $I$ band, while the faintest stars in the OGLE database have $I \approx 21$ mag. The typical photometric uncertainty for an individual measurement of a bright star ($I < 14$ mag) is about 0.005 mag.

**Table 1**
Newly Discovered Galactic Double Periodic Variables

| OGLE ID | $I$ (OGLE) (mag) | $V$ (OGLE) (mag) | $V - I$ (OGLE) (mag) | R.A. (J2000) | Decl. (J2000) |
|---|---|---|---|---|---|
| OGLE-BLG-ECL-022295 | 16.630 | 18.802 | 2.172 | 17:30:11.87 | −28:30:05.5 |
| OGLE-BLG-ECL-022845 | 16.137 | 19.294 | 3.157 | 17:30:30.80 | −30:50:26.1 |
| OGLE-BLG-ECL-026467 | 16.994 | 19.771 | 2.777 | 17:32:08.78 | −30:35:27.1 |
| OGLE-BLG-ECL-026993 | 16.778 | 19.291 | 2.513 | 17:32:20.45 | −28:00:25.8 |
| OGLE-BLG-ECL-029922 | 15.886 | 18.573 | 2.687 | 17:33:18.50 | −27:34:55.5 |
| OGLE-BLG-ECL-030962 | 18.252 | 21.177 | 2.925 | 17:33:35.56 | −28:33:39.9 |
| OGLE-BLG-ECL-058684 | 17.661 | 20.902 | 3.241 | 17:39:53.70 | −33:20:44.7 |
| OGLE-BLG-ECL-062890 | 16.534 | 18.699 | 2.165 | 17:40:45.16 | −37:41:37.3 |
| OGLE-BLG-ECL-072154 | 15.940 | 17.387 | 1.447 | 17:42:27.03 | −21:37:33.7 |
| OGLE-BLG-ECL-076504 | 16.789 | 18.727 | 1.938 | 17:43:10.65 | −34:29:42.5 |
| OGLE-BLG-ECL-083987 | 18.042 | 21.274 | 3.232 | 17:44:18.78 | −32:52:45.2 |
| OGLE-BLG-ECL-101353 | 17.118 | 19.719 | 2.601 | 17:46:41.42 | −21:52:56.1 |
| OGLE-BLG-ECL-116536 | 16.543 | 17.521 | 0.978 | 17:48:46.77 | −35:03:15.6 |
| OGLE-BLG-ECL-132549 | 17.108 | 20.168 | 3.060 | 17:50:48.55 | −30:40:20.0 |
| OGLE-BLG-ECL-137478 | 16.382 | 18.899 | 2.517 | 17:51:20.86 | −22:51:19.5 |
| OGLE-BLG-ECL-143267 | 16.613 | 19.050 | 2.437 | 17:51:53.66 | −32:52:22.0 |
| OGLE-BLG-ECL-184474 | 17.495 | 21.290 | 3.795 | 17:55:23.67 | −22:09:16.3 |
| OGLE-BLG-ECL-186594 | 16.253 | 17.663 | 1.410 | 17:55:34.15 | −32:49:34.2 |
| OGLE-BLG-ECL-189682 | 16.292 | 18.676 | 2.384 | 17:55:50.33 | −31:31:09.1 |
| OGLE-BLG-ECL-196052 | 16.097 | 18.809 | 2.712 | 17:56:21.93 | −20:31:35.7 |
| OGLE-BLG-ECL-198013 | 17.435 | 20.263 | 2.828 | 17:56:32.45 | −21:34:05.7 |
| OGLE-BLG-ECL-201730 | 15.586 | 16.467 | 0.881 | 17:56:51.04 | −35:09:18.6 |
| OGLE-BLG-ECL-210744 | 16.803 | ⋯ | ⋯ | 17:56:51.11 | −31:32:41.7 |
| OGLE-BLG-ECL-222537 | 15.737 | ⋯ | ⋯ | 17:58:38.36 | −34:25:27.4 |
| OGLE-BLG-ECL-271003 | 17.592 | 18.375 | 0.783 | 18:03:00.42 | −28:39:53.7 |
| OGLE-BLG-ECL-275974 | 15.866 | 17.637 | 1.771 | 18:03:27.99 | −27:43:43.1 |
| OGLE-BLG-ECL-279247 | 16.527 | 19.327 | 2.800 | 18:03:45.59 | −26:57:40.5 |
| OGLE-BLG-ECL-327914 | 15.308 | 17.584 | 2.276 | 18:08:32.17 | −25:29:53.2 |
| OGLE-BLG-ECL-328103 | 15.598 | 17.168 | 1.570 | 18:08:33.24 | −29:58:07.4 |
| OGLE-BLG-ECL-331967 | 15.786 | 17.309 | 1.523 | 18:08:59.52 | −26:36:48.5 |
| OGLE-BLG-ECL-369067 | 15.786 | 17.749 | 1.963 | 18:13:28.58 | −23:35:56.8 |
| OGLE-BLG-ELL-024487 | 14.353 | 15.186 | 0.833 | 18:18:18.63 | −28:07:26.6 |

## 3. Methodology

### 3.1. Long-period Search and Light-curve Disentangling

In order to find the secondary photometric variation, and consequently a new DPV candidate, we used several analysis tools. We subtracted the orbital periodic contribution provided by the catalog using Fourier analysis and worked with the code written by the late Zbigniew Kołaczkowski and explained by Mennickent et al. (2012). This method uses the frequency associated with the orbital period, and a fixed number of harmonics in order to recreate the contribution of luminosity variation due to the orbital motion of the system. After this procedure, the fitted Fourier approximated light curve is subtracted from the data in order to have only the residuals.

Those residuals were analyzed using the generalized Lomb–Scargle (GLS) periodogram (Zechmeister & Kürster 2009). This algorithm uses the principles of the Lomb (1976) and Scargle (1982) periodograms with some modifications, such as the addition of a displacement in the adjustment of the fit function and the consideration of measurement errors. Compared with the classical periodogram, it gives us more precise frequencies and a better determination of the amplitudes. The phase dispersion minimization algorithm (Stellingwerf 1978) allowed us to confirm the second periodicity. Considering the distribution of DPVs around the straight line $P_l = 33\ P_o$ (Mennickent 2017), we considered as DPV candidates those systems whose long periods are between the branches $P_l$max = $38.74\ P_o + 32.95$ days and $P_l$min = $38.74\ P_o - 192.05$ days, where most of the DPVs are expected to be found. To determine those constraints, we perform a linear regression to the previous 26 Galactic DPVs not passing through the origin of the parametric system (as the $P_l = 33\ P_o$ relation suggests). Later we moved the mentioned interpolation along the perpendicular direction of its position. Thus, at this first stage we excluded (likely very few) systems with extreme $P_l/P_o$ ratios. The $V, I$ colors are obtained from the OGLE online catalogs, except in Table 4 where they are from the analyzed time series. The periods reported here were calculated considering all photometric data available until 2019 October.

**Table 2**
Orbital Period, Long Period, and Period Ratio of Each DPV with Their Respective Errors

| Name | $P_{\rm orb}$ (days) | $P_{\rm long}$ (days) | $P_{\rm long}/P_{\rm orb}$ |
| --- | --- | --- | --- |
| OGLE-BLG-ECL-022295 | 14.678034 ± 0.000416 | 466.51 ± 1.40 | 31.78 ± 0.10 |
| OGLE-BLG-ECL-022845 | 19.715796 ± 0.001462 | 601.64 ± 2.51 | 30.52 ± 0.13 |
| OGLE-BLG-ECL-026467 | 20.78303 ± 0.000688 | 658.81 ± 4.33 | 31.70 ± 0.21 |
| OGLE-BLG-ECL-026993 | 14.761416 ± 0.000590 | 413.70 ± 1.25 | 28.03 ± 0.08 |
| OGLE-BLG-ECL-029922 | 8.107938 ± 0.000074 | 312.70 ± 0.75 | 38.57 ± 0.09 |
| OGLE-BLG-ECL-030962 | 14.883752 ± 0.000818 | 485.73 ± 2.81 | 32.63 ± 0.19 |
| OGLE-BLG-ECL-058684 | 21.70324 ± 0.001888 | 793.38 ± 4.11 | 36.56 ± 0.19 |
| OGLE-BLG-ECL-062890 | 16.66844 ± 0.006358 | 574.87 ± 18.58 | 34.49 ± 1.11 |
| OGLE-BLG-ECL-072154 | 15.04791 ± 0.000850 | 459.39 ± 1.51 | 30.53 ± 0.10 |
| OGLE-BLG-ECL-076504 | 1.598269 ± 0.000026 | 50.32 ± 0.02 | 31.48 ± 0.01 |
| OGLE-BLG-ECL-083987 | 11.017538 ± 0.000586 | 245.50 ± 1.02 | 22.28 ± 0.09 |
| OGLE-BLG-ECL-101353 | 12.997544 ± 0.001022 | 438.17 ± 8.51 | 33.71 ± 0.65 |
| OGLE-BLG-ECL-116536 | 7.712738 ± 0.000144 | 256.20 ± 0.29 | 33.22 ± 0.04 |
| OGLE-BLG-ECL-132549 | 12.17615 ± 0.000216 | 353.20 ± 0.45 | 29.01 ± 0.04 |
| OGLE-BLG-ECL-137478 | 21.46779 ± 0.003338 | 773.80 ± 6.85 | 36.04 ± 0.312 |
| OGLE-BLG-ECL-143267 | 16.803896 ± 0.000312 | 519.95 ± 0.68 | 30.94 ± 0.04 |
| OGLE-BLG-ECL-184474 | 18.64614 ± 0.002076 | 571.60 ± 4.82 | 30.66 ± 0.26 |
| OGLE-BLG-ECL-186594 | 1.592281 ± 0.000025 | 46.98 ± 0.04 | 29.50 ± 0.03 |
| OGLE-BLG-ECL-189682 | 17.222906 ± 0.000260 | 566.25 ± 0.72 | 32.88 ± 0.04 |
| OGLE-BLG-ECL-196052 | 25.744828 ± 0.004494 | 1091.06 ± 20.56 | 42.38 ± 0.80 |
| OGLE-BLG-ECL-198013 | 14.80092 ± 0.001594 | 521.81 ± 4.87 | 35.26 ± 0.33 |
| OGLE-BLG-ECL-201730 | 9.401168 ± 0.00043 | 291.63 ± 0.55 | 31.02 ± 0.06 |
| OGLE-BLG-ECL-210744 | 12.600516 ± 0.000106 | 299.57 ± 0.22 | 23.77 ± 0.02 |
| OGLE-BLG-ECL-222537 | 15.273438 ± 0.000528 | 419.40 ± 0.62 | 27.46 ± 0.04 |
| OGLE-BLG-ECL-271003 | 1.718797 ± 0.000018 | 37.84 ± 0.02 | 22.02 ± 0.01 |
| OGLE-BLG-ECL-275974 | 15.053288 ± 0.00022 | 451.22 ± 0.37 | 29.97 ± 0.02 |
| OGLE-BLG-ECL-279247 | 18.180322 ± 0.00014 | 526.07 ± 1.35 | 28.94 ± 0.07 |
| OGLE-BLG-ECL-327914 | 19.42795 ± 0.000588 | 681.60 ± 1.42 | 35.08 ± 0.07 |
| OGLE-BLG-ECL-328103 | 14.046314 ± 0.000382 | 414.58 ± 0.52 | 29.52 ± 0.04 |
| OGLE-BLG-ECL-331967 | 10.293578 ± 0.00008 | 374.77 ± 0.29 | 36.41 ± 0.03 |
| OGLE-BLG-ECL-369067 | 24.735716 ± 0.007972 | 827.92 ± 15.42 | 33.47 ± 0.62 |
| OGLE-BLG-ELL-024487 | 19.1256 ± 0.001004 | 605.06 ± 4.31 | 31.64 ± 0.23 |

## 4. Results

### 4.1. New DPVs

In Table 1, we provide the list of 32 new DPVs including coordinates, magnitudes, and $V-I$ colors. In Table 2 we provide the orbital and long periods and period ratio. This ratio has an average of 31.9 and a standard deviation of 4.4, calculated using the new and previously found galactic DPVs (58 in total). The new DPVs span an orbital period range of 1.6–26 days and a range of long-cycle lengths of 47–1144 days. We increased the number of known DPVs in the Milky Way by a factor of 2.

### 4.2. Gaia Identifications, Reddening, and Extinction

The location of the newly discovered DPVs in the direction of the Galactic bulge (see Figure 1) causes the light from the objects to be strongly reddened by unevenly distributed interstellar matter (Nataf et al. 2013; Iwanek et al. 2019).

We cross-matched our list of new DPVs with Gaia DR2 within a radius of 0.4 arcsec. We found 31 out of 32 objects, missing only OGLE-BLG-ECL-030962. To obtain reddening and extinction we made several assumptions: (1) If the parallax is negative in Gaia DR2 we assumed that the object is far away, and we put "NA" in the parallax and error for this object, and we assumed that this object is in the bulge, at a distance of 8.0 kpc. (2) If parallax $p_i$ is measured not so accurately ($\sigma_{p_i}/p_i > 0.2$), we also put "NA" in the parallax and error for

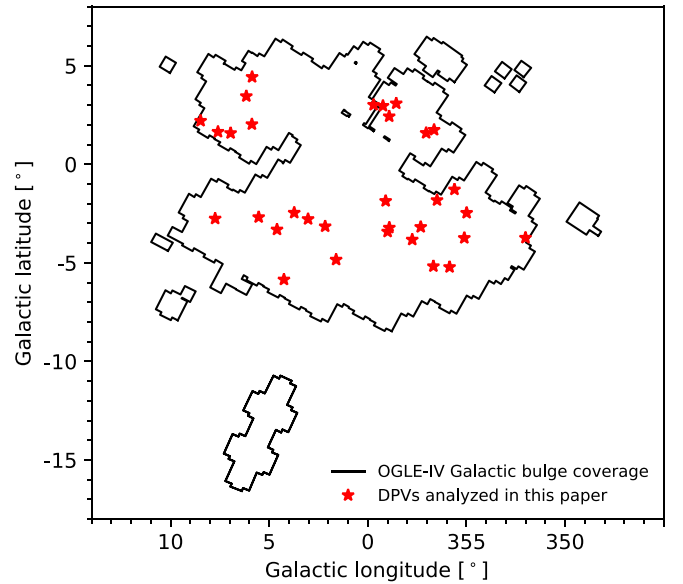

**Figure 1.** Distribution of 32 newly discovered DPVs systems (marked with red asterisks) in the sky. The black solid line presents the OGLE-IV Galactic bulge coverage.

this object, and again we have assumed a distance to this object of 8.0 kpc. (3) For the object that is missing in Gaia DR2, viz. OGLE-BLG-ECL-030962, we put also "NA" in the parallax and error columns, and a distance of 8.0 kpc. (4) Only three

Table 3
Supplementary Data for New Double Periodic Variables

| Object | Gal. Long. (deg) | Gal. Lat. (deg) | $p$[a] (mas) | $p$ Error (mas) | $d$[b] (kpc) | $A_V$[c] (mag) | $A_I$[c] (mag) | $E(V-I)$[c] mag | $A_V$[d] (mag) | $A_I$[d] (mag) | $E(V-I)$[d] (mag) |
|---|---|---|---|---|---|---|---|---|---|---|---|
| OGLE-BLG-ECL-022295 | 358.557628 | 3.089058 | −0.2414 | 0.4955 | 8.000 | NA | NA | NA | 3.591 | 1.971 | 1.620 |
| OGLE-BLG-ECL-022845 | 356.638929 | 1.746973 | 0.1505 | 0.13462 | 8.000 | NA | NA | NA | 5.373 | 2.949 | 2.424 |
| OGLE-BLG-ECL-026467 | 357.040308 | 1.590201 | 0.069 | 0.4050 | 8.000 | NA | NA | NA | 4.358 | 2.392 | 1.966 |
| OGLE-BLG-ECL-026993 | 359.230521 | 2.964263 | 0.1195 | 0.3709 | 8.000 | NA | NA | NA | 4.042 | 2.219 | 1.824 |
| OGLE-BLG-ECL-029922 | 359.704265 | 3.015798 | 0.4717 | 0.1379 | 8.000 | NA | NA | NA | 4.599 | 2.525 | 2.075 |
| OGLE-BLG-ECL-030962 | 358.915140 | 2.431787 | NA | NA | 8.000 | NA | NA | NA | 4.582 | 2.515 | 2.067 |
| OGLE-BLG-ECL-058684 | 355.605523 | −1.278726 | 0.2229 | 0.3750 | 8.000 | NA | NA | NA | 5.115 | 2.807 | 2.307 |
| OGLE-BLG-ECL-062890 | 352.005548 | −3.728611 | NA | NA | 8.000 | NA | NA | NA | 3.198 | 1.756 | 1.443 |
| OGLE-BLG-ECL-072154 | 5.864432 | 4.430295 | 1.668 | 0.2928 | 0.599 | 2.359 | 1.296 | 1.063 | 0.800 | 0.439 | 0.361 |
| OGLE-BLG-ECL-076504 | 354.988326 | −2.463740 | −0.0921 | 0.1939 | 8.000 | 3.179 | 1.743 | 1.436 | 2.990 | 1.641 | 1.349 |
| OGLE-BLG-ECL-083987 | 356.489859 | −1.819114 | NA | NA | 8.000 | NA | NA | NA | 5.794 | 3.180 | 2.614 |
| OGLE-BLG-ECL-101353 | 6.157766 | 3.456031 | 0.0847 | 0.2451 | 8.000 | NA | NA | NA | 3.343 | 1.835 | 1.508 |
| OGLE-BLG-ECL-116536 | 355.106546 | −3.737345 | 0.6732 | 0.2998 | 8.000 | 1.932 | 1.076 | 0.856 | 1.981 | 1.087 | 0.894 |
| OGLE-BLG-ECL-132549 | 359.094035 | −1.860209 | NA | NA | 8.000 | 4.527 | 2.472 | 2.055 | 4.176 | 2.292 | 1.884 |
| OGLE-BLG-ECL-137478 | 5.875026 | 2.032450 | −0.0043 | 0.1786 | 8.000 | 3.794 | 2.089 | 1.705 | 3.361 | 1.845 | 1.516 |
| OGLE-BLG-ECL-143267 | 357.316377 | −3.179985 | 0.3756 | 0.2267 | 8.000 | 3.202 | 1.780 | 1.422 | 2.954 | 1.621 | 1.332 |
| OGLE-BLG-ECL-184474 | 6.953516 | 1.582069 | −0.1959 | 0.3676 | 8.000 | NA | NA | NA | 4.530 | 2.486 | 2.043 |
| OGLE-BLG-ECL-186594 | 357.746645 | −3.822959 | 0.0306 | 0.1319 | 8.000 | 2.371 | 1.302 | 1.069 | 2.326 | 1.277 | 1.049 |
| OGLE-BLG-ECL-189682 | 358.908536 | −3.217529 | 0.1511 | 0.1980 | 8.000 | 3.071 | 1.659 | 1.412 | 2.583 | 1.418 | 1.165 |
| OGLE-BLG-ECL-196052 | 8.475284 | 2.205042 | −0.0218 | 0.1831 | 8.000 | NA | NA | NA | 3.520 | 1.932 | 1.588 |
| OGLE-BLG-ECL-198013 | 7.594280 | 1.646958 | 0.5006 | 0.4056 | 8.000 | NA | NA | NA | 5.215 | 2.862 | 2.353 |
| OGLE-BLG-ECL-201730 | 355.853858 | −5.213840 | −0.01184 | 0.2476 | 8.000 | 1.32 | 0.736 | 0.584 | 1.566 | 0.859 | 0.706 |
| OGLE-BLG-ECL-210744 | 358.994402 | −3.417298 | NA | NA | 8.000 | 3.019 | 1.662 | 1.357 | 2.715 | 1.490 | 1.225 |
| OGLE-BLG-ECL-222537 | 356.674097 | −5.171711 | 0.1177 | 0.1708 | 8.000 | 1.204 | 0.651 | 0.553 | 1.318 | 0.723 | 0.595 |
| OGLE-BLG-ECL-271003 | 2.160899 | −3.148999 | 0.1610 | 0.5863 | 8.000 | 1.836 | 0.992 | 0.844 | 1.530 | 0.840 | 0.690 |
| OGLE-BLG-ECL-275974 | 3.027708 | −2.778162 | 0.2200 | 0.1463 | 8.000 | 2.082 | 1.109 | 0.973 | 1.972 | 1.083 | 0.890 |
| OGLE-BLG-ECL-279247 | 3.729224 | −2.458737 | −0.6954 | 0.2330 | 8.000 | 3.165 | 1.698 | 1.467 | 2.494 | 1.369 | 1.125 |
| OGLE-BLG-ECL-327914 | 5.530497 | −2.682171 | NA | NA | 8.000 | 2.747 | 1.499 | 1.248 | 2.760 | 1.515 | 1.245 |
| OGLE-BLG-ECL-328103 | 1.605719 | −4.838699 | NA | NA | 8.000 | 1.396 | 0.768 | 0.628 | 1.495 | 0.820 | 0.674 |
| OGLE-BLG-ECL-331967 | 4.601795 | −3.310174 | 0.2199 | 0.1019 | 8.000 | 2.366 | 1.263 | 1.103 | 2.087 | 1.146 | 0.942 |
| OGLE-BLG-ECL-369067 | 7.738126 | −2.755571 | −0.0521 | 0.1387 | 8.000 | NA | NA | NA | 3.335 | 1.830 | 1.504 |
| OGLE-BLG-ELL-024487 | 4.249912 | −5.847587 | 0.1452 | 0.0564 | 8.000 | NA | NA | NA | 1.106 | 0.607 | 0.499 |

**Notes.** See the text for details.
[a] Parallaxes from GDR2.
[b] Distances calculated using the GDR2 parallax.
[c] Extinction and reddening from Nataf et al. (2013).
[d] Extinction and reddening from mwdust.

objects have nonnegative and accurate parallax measurements, so for them we have calculated distances resulting from the parallax. As a result, we identified 31 DPVs as members of the Galactic bulge, and 1 DPV that is located closer to the Sun, and does not belong to the Galactic bulge, but it is projected on the field of view of the bulge, viz. OGLE-BLG-ECL-072154, located at 0.599 kpc.

In the first step we obtained reddening and extinction using Nataf's maps. These maps are made using red clump stars from the bulge, using OGLE-III data. Using these maps we automatically make the assumption that the object for which we measure extinction and reddening is located in the bulge, behind all the dust. These maps are good enough for distant stars (which are located more or less in the bulge). The "NA" label in columns 7–9 (Nataf's reddening and extinction) means that toward these directions there are no measurements available.

We calculated dereddened colors and magnitudes defined as

$$I_0 = I - A_I, \quad (1)$$

$$(V - I)_0 = (V - I) - E(V - I), \quad (2)$$

where $A_I$ and $E(V - I)$ are the $I$-band extinction and the color excess, respectively. Table 3 provides a compilation of our results about Gaia identification, extinction, and reddening. Additional information about the analyzed light curves is given in Table 4.

## 5. Discussion

In Figure 2 we show the correlation between orbital and long periods for previously known Galactic DPVs supplemented with new discoveries described in this paper. It is clear that the new systems follow the same general tendency, in the sense of a monotonic increase of the orbital period with a long-cycle length. Most of the new systems have periods longer than 10 days. MNIC V99 reported by Nikolay Mishevskiy in VSX, has a period ratio of 21.67, similar to $\beta$ Lyrae (21.25). It is a member of the open cluster NGC 146, and was classified as a classical Be star of spectral type B3V by Mathew & Subramaniam (2011). The period distributions are shown in Figure 3. The distribution of dereddened $(V - I)_0$ colors and magnitudes are given in Figure 4. Most of the systems have

Table 4
Observational Data for Each DPV[a]

| DPV OGLE-BGL- | Filter | $N_{data}$ | $T_i$ | $T_f$ | $\bar{m}$ ag | $\sigma$ mag |
|---|---|---|---|---|---|---|
| ECL-022295 | I | 1172 | 5265.81414 | 8763.57341 | 16.695 | 0.067 |
| $V - I = 2.172 \pm 0.019$ | V | 65 | 5267.80384 | 8726.61207 | 18.867 | 0.080 |
| ECL-022845 | I | 922 | 5309.86179 | 8783.52539 | 16.277 | 0.212 |
| $V - I = 3.187 \pm 0.029$ | V | 45 | 5641.85396 | 8728.57093 | 19.464 | 0.419 |
| ECL-026467 | I | 942 | 5307.81611 | 8783.52539 | 17.059 | 0.060 |
| $V - I = 2.789 \pm 0.037$ | V | 45 | 5641.85396 | 8728.57093 | 19.848 | 0.065 |
| ECL-026993 | I | 1034 | 5376.66303 | 8763.57341 | 16.920 | 0.190 |
| $V - I = 2.574 \pm 0.033$ | V | 60 | 5267.80384 | 8726.61207 | 19.493 | 0.332 |
| ECL-029922 | I | 7212 | 5265.81251 | 8783.51337 | 15.920 | 0.035 |
| $V - I = 2.689 \pm 0.014$ | V | 211 | 5267.80162 | 8763.54658 | 18.609 | 0.037 |
| ECL-030962 | I | 1149 | 5288.86318 | 8783.52846 | 18.401 | 0.247 |
| $V - I = 3.237 \pm 0.176$ | V | 44 | 5288.84266 | 8741.59303 | 21.638 | 0.376 |
| ECL-058684 | I | 744 | 5306.76435 | 8762.52513 | 17.800 | 0.220 |
| $V - I = 3.426 \pm 0.157$ | V | 28 | 5740.70369 | 8695.76659 | 21.227 | 0.274 |
| ECL-062890 | I | 172 | 5385.58089 | 6737.82676 | 16.687 | 0.174 |
| $V - I = 2.239 \pm 0.027$ | V | 6 | 5694.72823 | 6534.56085 | 18.926 | 0.105 |
| ECL-072154 | I | 549 | 2124.50928 | 7817.83566 | 16.048 | 0.202 |
| $V - I = 1.451 \pm 0.010$ | V | 29 | 4757.5214 | 7269.53372 | 17.500 | 0.079 |
| ECL-076504 | I | 1820 | 2116.77407 | 8750.60203 | 16.810 | 0.136 |
| $V - I = 1.952 \pm 0.023$ | V | 59 | 3480.8138 | 8716.54655 | 18.762 | 0.068 |
| ECL-083987 | I | 926 | 5378.57053 | 8762.51912 | 18.217 | 0.257 |
| $V - I = 3.252 \pm 0.142$ | V | 53 | 5642.81836 | 8716.54874 | 21.469 | 0.308 |
| ECL-101353 | I | 659 | 5265.84296 | 7683.50095 | 17.261 | 0.187 |
| $V - I = 2.696 \pm 0.050$ | V | 31 | 5275.88529 | 7284.50657 | 19.957 | 0.378 |
| ECL-116536 | I | 1959 | 2115.78469 | 8763.57847 | 16.594 | 0.167 |
| $V - I = 1.043 \pm 0.018$ | V | 79 | 3478.8164 | 8718.49908 | 17.637 | 0.399 |
| ECL-132549 | I | 8040 | 2125.56122 | 8787.51017 | 17.172 | 0.131 |
| $V - I = 3.140 \pm 0.079$ | V | 169 | 3476.85723 | 8728.5651 | 20.312 | 0.219 |
| ECL-137478 | I | 536 | 5265.85454 | 7660.52945 | 16.503 | 0.188 |
| $V - I = 2.532 \pm 0.035$ | V | 24 | 5276.85882 | 7226.66062 | 19.035 | 0.102 |
| ECL-143267 | I | 3161 | 551.8965 | 8783.55044 | 16.753 | 0.196 |
| $V - I = 2.469 \pm 0.027$ | V | 174 | 1021.77345 | 8725.4965 | 19.222 | 0.306 |
| ECL-184474 | I | 397 | 5265.86046 | 7667.49853 | 17.648 | 0.180 |
| $V - I = 3.920 \pm 0.186$ | V | 11 | 5276.86706 | 7198.70108 | 21.568 | 0.342 |
| ECL-186594 | I | 1934 | 5260.86866 | 8755.59956 | 16.337 | 0.100 |
| $V - I = 1.421 \pm 0.012$ | V | 156 | 5266.85076 | 8722.52336 | 17.758 | 0.124 |
| ECL-189682 | I | 4097 | 2125.53977 | 8783.5535 | 16.413 | 0.143 |
| $V - I = 2.419 \pm 0.021$ | V | 161 | 3476.83803 | 8724.57564 | 18.832 | 0.245 |
| ECL-19605 | I | 389 | 5265.85896 | 7666.5734 | 16.227 | 0.152 |
| $V - I = 2.669 \pm 0.019$ | V | 13 | 5276.86496 | 7197.73354 | 18.896 | 0.100 |
| ECL-198013 | I | 385 | 5265.85896 | 7666.5734 | 17.585 | 0.186 |
| $V - I = 2.915 \pm 0.069$ | V | 13 | 5276.86496 | 7197.73354 | 20.500 | 0.410 |
| ECL-201730 | I | 768 | 2115.7781 | 7673.55067 | 15.719 | 0.255 |
| $V - I = 0.891 \pm 0.009$ | V | 40 | 4686.58622 | 7252.5753 | 16.610 | 0.309 |
| ECL-210744 | I | 19179 | 2072.6039 | 8787.51326 | 16.969 | 0.283 |
| ECL-222537 | I | 524 | 2116.78882 | 8617.9366 | 15.880 | 0.197 |
| $V - I = 1.219 \pm 0.010$ | V | 3 | 3480.83224 | 4916.88887 | 17.099 | 0.054 |
| ECL-271003 | I | 15831 | 551.84063 | 8787.51786 | 17.602 | 0.139 |
| $V - I = 0.817 \pm 0.025$ | V | 247 | 1300.6775 | 8742.58799 | 18.419 | 0.154 |

Table 4
(Continued)

| DPV OGLE-BGL- | Filter | $N_{data}$ | $T_i$ | $T_f$ | $\bar{m}$ ag | $\sigma$ mag |
|---|---|---|---|---|---|---|
| ECL-275974 | I | 8864 | 2127.48967 | 8787.51634 | 16.013 | 0.214 |
| $V - I = 1.801 \pm 0.011$ | V | 217 | 3476.79531 | 8742.58592 | 17.813 | 0.341 |
| ECL-279247 | I | 8885 | 2127.51889 | 8787.51634 | 16.619 | 0.080 |
| $V - I = 2.814 \pm 0.029$ | V | 217 | 3476.81034 | 8742.58592 | 19.433 | 0.088 |
| ECL-327914 | I | 2341 | 2127.60225 | 8763.5681 | 15.395 | 0.104 |
| $V - I = 2.307 \pm 0.009$ | V | 187 | 4705.53655 | 8723.52314 | 17.702 | 0.191 |
| ECL-328103 | I | 1135 | 2125.61109 | 8782.55958 | 15.724 | 0.181 |
| $V - I = 1.602 \pm 0.010$ | V | 55 | 4703.51861 | 8724.5801 | 17.326 | 0.307 |
| ECL-331967 | I | 3480 | 551.80801 | 8763.5696 | 15.922 | 0.169 |
| $V - I = 1.575 \pm 0.011$ | V | 170 | 607.91592 | 8725.56589 | 17.496 | 0.267 |
| ECL-369067 | I | 289 | 5278.90253 | 7684.52047 | 15.904 | 0.282 |
| $V - I = 1.927 \pm 0.014$ | V | 9 | 5662.82138 | 7221.64591 | 17.831 | 0.053 |
| ELL-024487 | I | 475 | 2463.60617 | 7672.54903 | 14.402 | 0.042 |

**Note.**
[a] Initial and final dates of time series are given (HJD 2,450,000). $V, I$ is the uncorrected color obtained from the average magnitudes.

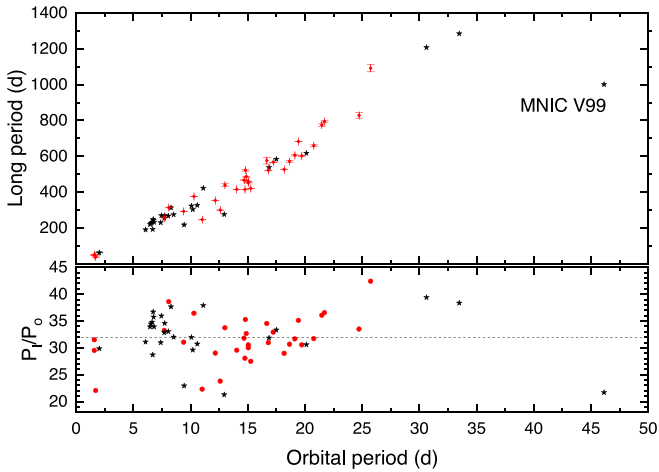

**Figure 2.** Long period and period ratio vs. orbital period for Galactic DPVs. The new DPVs reported in this paper are shown by dots while the rest, marked with asterisks, are listed in Mennickent (2017), Rosales & Mennickent (2018, 2019), and the VSX database (https://www.aavso.org/vsx/). MNIC V99, reported by Nikolay Mishevskiy in VSX, has a period ratio of 21.67, similar to $\beta$ Lyrae (21.25). The average $P_l/P_o = 31.9$ is indicated.

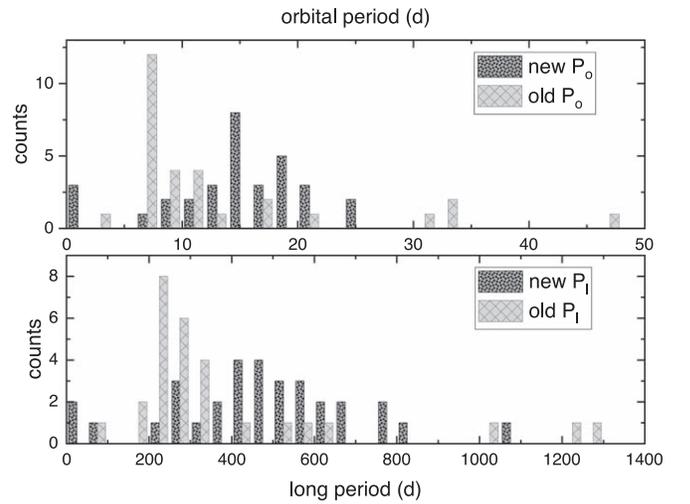

**Figure 3.** Period distribution for Galactic DPVs.

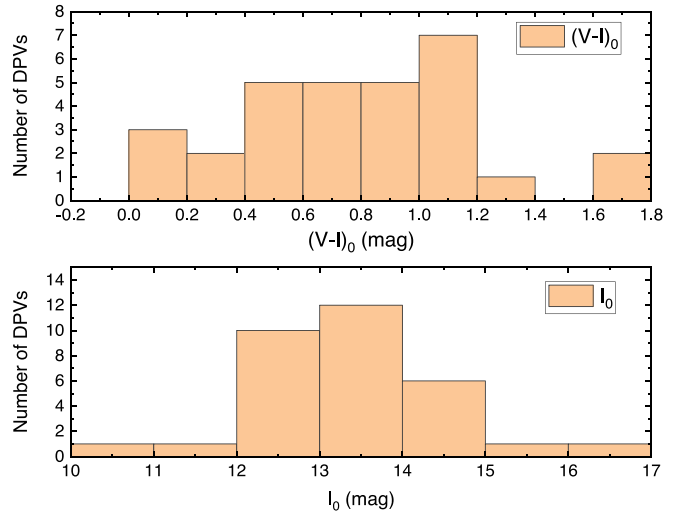

**Figure 4.** Dereddened color and magnitude distributions for Galactic DPVs.

brightness $I_0$ between 12 and 15 mag, while the $(V - I)_0$ colors span a range between 0.0 and 1.8.

Some examples of light curves are given in Figure 5.[5] They show typical orbital eclipsing light curves of $\beta$ Lyrae type, i.e., with rounded light curve segments between eclipses, as a result of the proximity among the stars and the gravitational deformation of the less massive component. In Figure 6 we show the color–magnitude diagram constructed with dereddened colors and magnitudes along with PARSEC-COLIBRI stellar isochrones for solar metallicity including the TP-AGB phase (Marigo et al. 2017). Considering that the stars in the Galactic bulge of the Milky Way have a broad range of metallicity, $-3.0 < [Fe/H] < 1$ dex (Ness & Freeman 2016),

---
[5] The full data sample of discovered DPVs is available at http://www.astrouw.edu.pl/ogle/ogle4/OCVS/blg/dpv/.

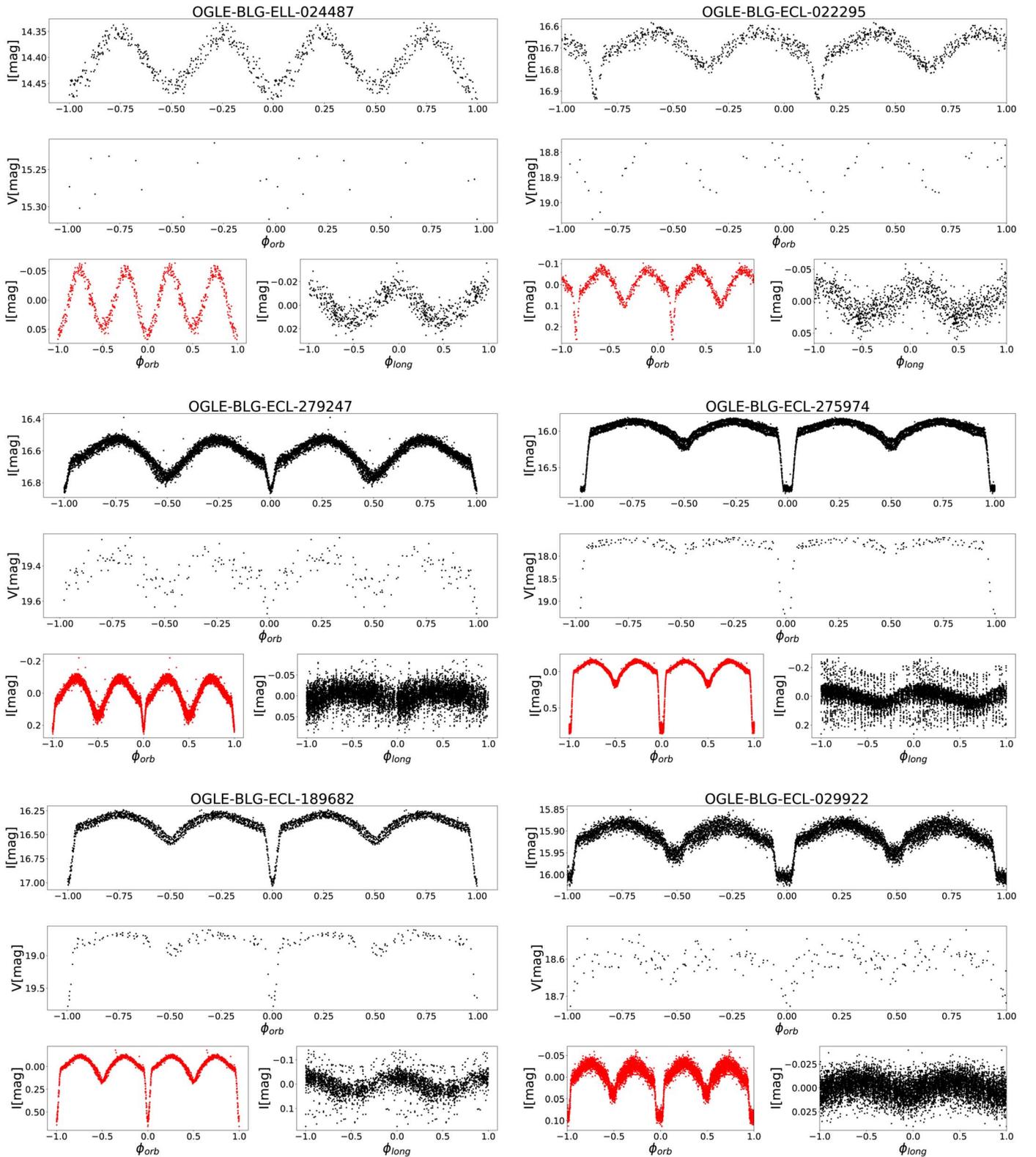

**Figure 5.** Examples of light curves. Upper: observed light curve. Bottom left: disentangled orbital light curves. Bottom right: disentangled long-cycle light curve. Visit http://www.astrouw.edu.pl/ogle/ogle4/OCVS/blg/dpv/ for the complete data sample.

we compared our DPVs with isochrones of a representative population of solar metallicity single stars, expecting to infer only general highlights for our DPV population. Keeping in mind that mass transfer usually modifies the evolution of the close binary components, it is apparent that the newly discovered DPV population is located far from the main-sequence placement. This can be explained if the DPV $V - I$ colors are dominated by the cool mass-transferring giant star. The most extreme red colors in OGLE-BLG-ECL-184474 and OGLE-BLG-ECL-279247 might be explained by light

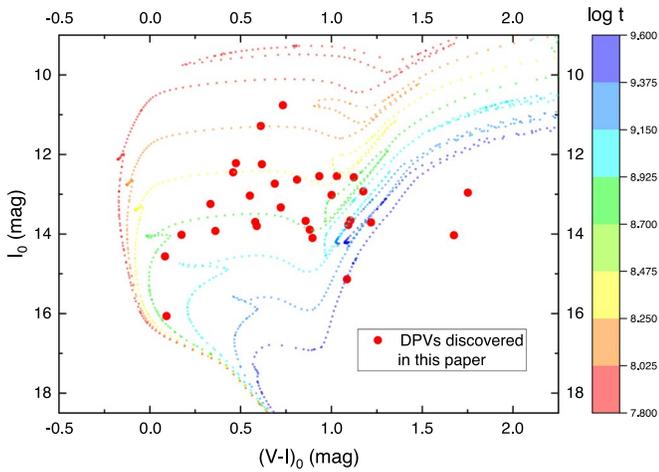

**Figure 6.** Dereddened color–magnitude diagram for the new DPVs marked with red dots. We plot PARSEC-COLIBRI stellar isochrones for solar metallicity including the TP-AGB phase (Marigo et al. 2017). Isochrones are marked with dotted lines for different values of log t (Gyr).

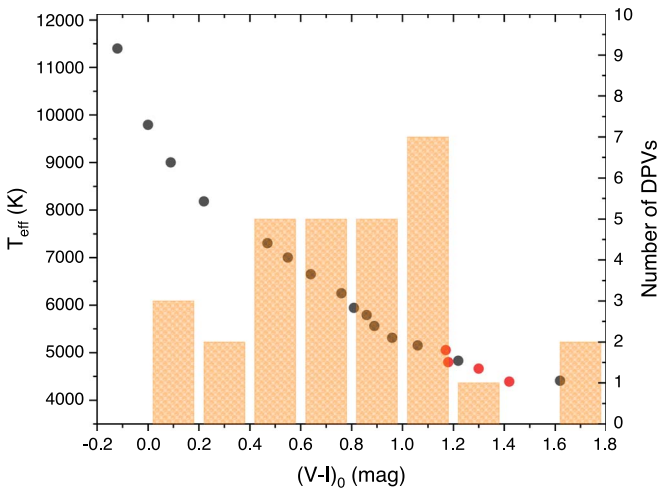

**Figure 7.** Histogram of colors of new DPVs compared with colors of dwarfs (black dots) and giants (red dots) of different effective temperatures from Cox (2000).

emission from cool circumstellar material, a large accretion disk for instance.

We compare DPV colors with those presented in single giant and dwarf stars (Figure 7). If the donor dominates the system light, then their temperatures are in the range of 11,000–4000 K, corresponding to spectral types ∼B8–K8. It is likely that the true range is in fact narrower, since infrared excess due to circumstellar matter could explain part of the redder colors, while the bluer ones might be influenced by hot and luminous primary components.

## 6. Summary

We have found 32 new binaries of the DPV type, doubling the number of known DPVs in the Milky Way. Most of the new DPVs are located in the Galactic bulge. We determined orbital periods and long-cycle periods for all of them and provide coordinates, color excess, and $I$, $V − I$ information. The $I_0$ versus $(V − I)_0$ distribution suggests a binary population where the light is dominated by cool giant secondaries and circumstellar matter. Some of the new systems show total eclipses, and are candidates for precise modeling in future investigations.

G.R.G. acknowledges the financial support of the National Science Centre, Poland (NCN) under the project No. 2017/26/A/ST9/00446. R.E.M. acknowledges support by the BASAL Centro de Astrofísica y Tecnologías Afines (CATA) PFB–06/2007 and FONDECYT 1190621. P.I. acknowledges the financial support of the Polish National Science Centre through the PRELUDIUM grant No. 2019/35/N/ST9/02474. The OGLE project has received funding from Polish National Science Centre grant MAESTRO No. 2014/14/A/ST9/00121. N. A.-D. acknowledges the support of FONDECYT project 3180063. J.G. acknowledges ANID project 21202285.

### ORCID iDs

Gonzalo Rojas García  https://orcid.org/0000-0001-9266-3036
Ronald Mennickent  https://orcid.org/0000-0002-6245-0264
P. Iwanek  https://orcid.org/0000-0002-6212-7221
I. Soszyński  https://orcid.org/0000-0002-7777-0842